\documentclass[aps,prb,twocolumn,showpacs,superscriptaddress,groupedaddress,nofootinbib]{revtex4}  
\usepackage{bm}
\usepackage{mathrsfs}
\usepackage{amsmath}
\usepackage{amssymb}
\usepackage{graphicx}
\usepackage{amsfonts}
\usepackage{cancel}
\usepackage{amsthm}
\usepackage{color}
\usepackage{dcolumn}
\usepackage{txfonts}
\usepackage{subfigure}
\begin{document}


\title{Giant Faraday rotation induced by Berry phase in bilayer graphene under strong terahertz fields}

\author{ Fan Yang }
\affiliation{Department of Physics, The Chinese University of Hong Kong, Shatin, N.T., Hong Kong, China}

\author{ Xiaodong Xu }
\affiliation{Department of Physics, University of Washington, Seattle, Washington 98195, USA}

\author{ Ren-Bao Liu }
\email{rbliu@phy.cuhk.edu.hk}
\affiliation{Department of Physics, The Chinese University of Hong Kong, Shatin, N.T., Hong Kong, China}
\affiliation{Institute of Theoretical Physics, The Chinese University of Hong Kong, Shatin, N.T., Hong Kong, China}


\date{\today}

\begin{abstract}

High-order terahertz (THz) sideband generation (HSG) in semiconductors is a phenomenon with physics similar to high-order harmonic generation but in a much lower frequency regime. It was found that the electron-hole pairs excited by a weak optical laser can accumulate Berry phases along a cyclic path under the driving of a strong THz field. The Berry phases appear as the Faraday rotation angles of the emission signal under short-pulse excitation in monolayer MoS$_2$. In this paper, the theory of Berry phase in THz extreme nonlinear optics is applied to biased bilayer graphene with Bernal stacking, which has similar Bloch band features and optical properties to the monolayer MoS$_2$, such as time-reversal related valleys and valley contrasting optical selection rules. The bilayer graphene has much larger Berry curvature than monolayer MoS$_2$, which leads to a giant Faraday rotation of the optical emission ($\sim$ 1 rad for a THz field with frequency 1 THz and strength 8 kV/cm). This provides opportunities to use bilayer graphene and low-power THz lasers for ultrafast electro-optical devices.

\end{abstract}
\maketitle

\section{Introduction}

In semiconductors irradiated by a strong terahertz (THz) laser of frequency $\omega$, the electron-hole pairs excited by a weak optical laser of frequency $\Omega$ are driven into large amplitude oscillations. They subsequently acquire kinetic energies much greater than the THz photon energy during the oscillations, which leads to optical emission at frequencies $\Omega \pm 2N \omega$, where $N$ is an integer.~\cite{HSG_RBL,Zaks_Liu} This high-order THz-sideband generation (HSG) in semiconductors is a generalization of the high-order harmonic generation (HHG)~\cite{HHG_Krause,HHG_Corkum,HHG_QT,HHG_Phytoday} to the THz frequency regime. The sidebands in the frequency domain indicate THz modulation in the time domain, which has potential applications in Tbit/s optical communications, trillion-FPS optical imaging and wideband optical multiplexers.

The core physics of HHG and HSG is captured by the quantum trajectory theory.~\cite{HSG_RBL,HHG_QT} When the electrons (or electron-hole pairs) are driven by the strong laser field, their oscillation amplitudes are much larger than the wavepacket diffusion range. Therefore the quantum evolution of the electrons (or electron-hole pairs) is well described by quantum trajectories that satisfy the stationary phase condition (i.e., the least action condition in classical mechanics) in the formalism of Dirac-Feynmann path integrals.~\cite{HSG_RBL,HHG_QT}

A fundamental difference between HHG and HSG is that the electron-hole pair in semiconductors can have nontrivial Bloch states.~\cite{YF_MoS2} As a result, when the eletron (or hole) is driven by the THz field ${\mathbf F}\left(t\right)$ in the conduction (or valence) bands, not only does the quasi-momentum change according to $\dot {{\mathbf k}}=-e{\mathbf F}\left(t\right)$,~\cite{Solid_AM} but also does its Bloch wavefunction evolve with $\mathbf k$. Since the THz frequency is much smaller than the band gaps of semiconductors, the evolution is adiabatic (i.e., no interband tunneling is induced by the THz field). This adiabatic evolution along the quantum trajectory in semiconductors can accumulate a geometrical phase, in particular a Berry phase for a cyclic evolution.~\cite{Berry} A rich structures of the Bloch states in condensed matter systems (such as in topological insulators~\cite{SCZhang,Kane}) thus leads to a variety of phase effects in extreme nonlinear optics.

An additional important difference between the HHG and HSG physics is that the electron-hole pairs are elementary optical excitations in solids. Therefore, the quantum trajectories in semiconductors, unlike those in atoms, can be triggered by lasers on demand at designed frequencies~\cite{XTXie} or times (relative to the THz field oscillation). This excitation at will provides a great deal of controllability and flexibility for studying the quantum trajectories in extreme nonlinear optics.

In Ref. \citenum{YF_MoS2}, the Berry phase dependent quantum trajectory theory was applied to monolayer MoS$_2$ with a band gap in the visible wavelength regime. The optical emission delayed by integer multiples of the THz period after the pulse excitation acquires a Faraday rotation. The rotation angle was shown to be exactly the Berry phase of the quantum trajectory ($\sim$ 0.01 rad for a THz field with frequency 1 THz and strength 8 kV/cm). That result provides new opportunities to utilize THz extreme nonlinear optics of thin film materials for ultrafast electro-optical devices.

\begin{figure}
\begin{center}
\includegraphics[width=\columnwidth]{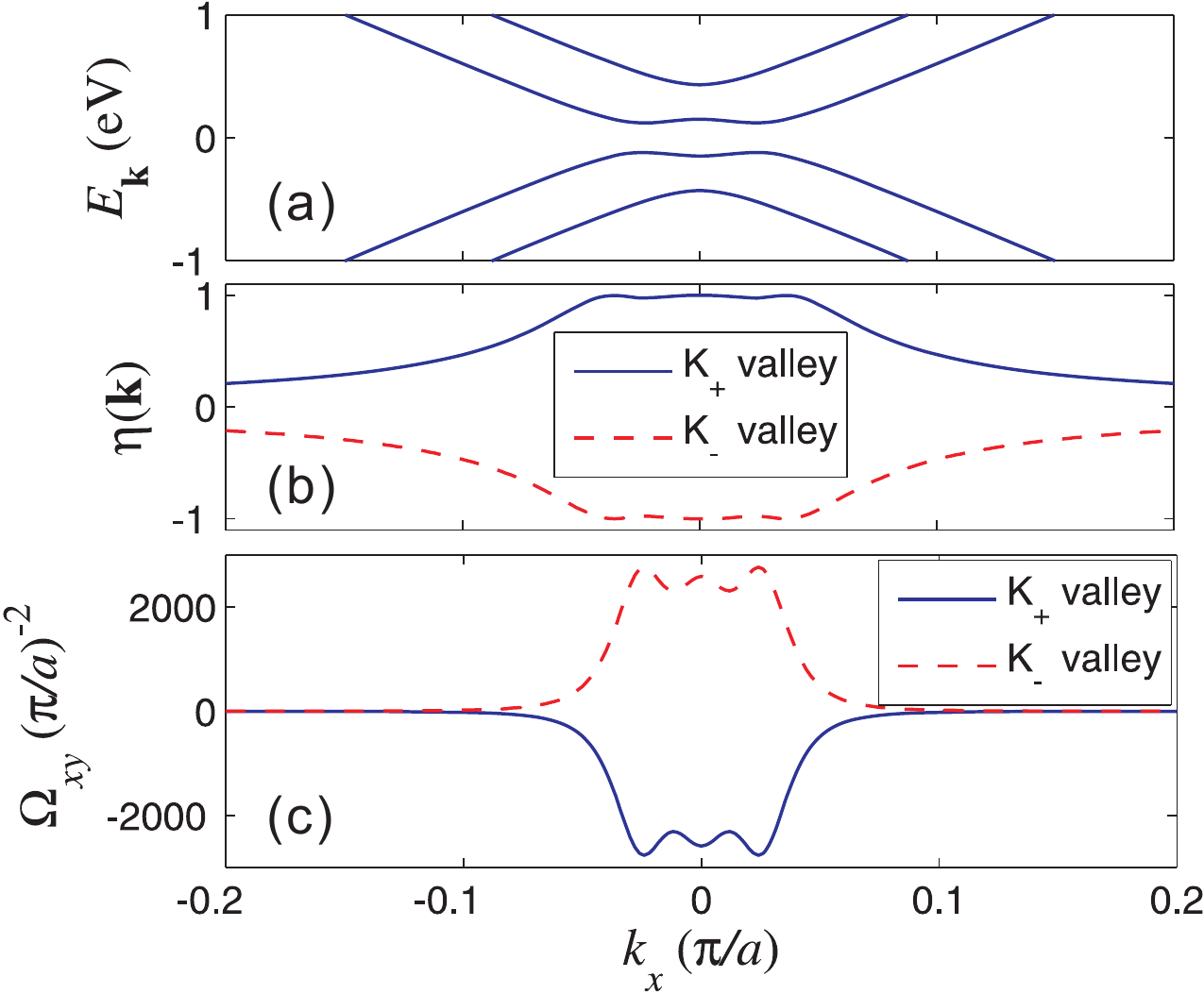}
\end{center}
\caption{(color online). Energy bands, optical slection rules, and Berry curvatures near the Dirac points of a biased bilayer graphene. (a) shows the energy spectrum at the $K_{\pm}$ valleys, where $\mathbf k$ is expanded around the respective Dirac points and $k_y=0$. (b) shows the degrees of circular polarization for the interband dipole moment (defined in Eq. (\ref{deg_polar})), where we see the valley contrasting optical selection rules near the two Dirac points. (c) shows the combined Berry curvature of the bottom conduction band and the top valence band. The solid (dashed) line gives the values in valley $K_+$ ($K_-$). The interlayer bias is chosen as $2\Delta=0.3$ eV. } \label{Optical_sel}
\end{figure}

In this paper, we consider the transient optical response of Bernal stacked bilayer graphene excited by a weak optical pulse and a strong THz field. The bilayer graphene with an interlayer gate bias~\cite{BilayerG,Graphene_RMP,Bilayer_Tunable,Bilayer_ARPES} has similar Bloch band features and optical properties as the monolayer MoS$_2$ but with a smaller band gap tunable up to $\sim250$ meV. Its conduction and valence-band edges are located at the corners of the 2D hexagonal Brillouin zone, and similar to MoS$_2$, the optical interband transitions at the two time-reversal (TR) related valleys have nearly perfect but opposite polarization selection rules.~\cite{MOS2,YAO_selection} These features make bilayer graphene an excellent candidate for the observation of the Berry phases of quantum trajectories. In bilayer graphene without interlayer bias, the conduction (valence) state acquires a $\pm 2\pi$ Berry phase (pseudospin winding number being $\pm 2$) along any closed path around the Dirac points,~\cite{Graphene_RMP,Pseudo_widing} and the Berry curvature is completely concentrated on the singular monopole at the Dirac point. In biased bilayer graphene that has an energy gap opened at the Dirac points, the Berry curvature becomes non-singular while its distribution is still concentrated in a small region in the vicinity of the Dirac points (since the Berry curvature of the region far from the Dirac point is zero as in the unbiased case) (see Fig.~\ref{Optical_sel}). Thus the Berry curvature in bilayer graphene is much (about 100 times) larger than in monolayer MoS$_2$. This leads to a much larger Berry phase of a quantum trajectory and in turn giant Faraday rotation. In other words, bilayer graphene requires a much weaker THz laser to induce an observable Faraday rotation, and serves as a better material for ultrafast electro-optical devices than MoS$_2$. Besides, the band gap at the Dirac points is tunable by an applied gate bias between zero and midinfrared energies,~\cite{Bilayer_Tunable,Bilayer_ARPES} which offers outstanding controllability.

This paper is organized as follows. In Sec. \ref{genernal}, we present a general theory of optical response in semiconductors under a strong THz field, with the Berry phase effects included. The optical response under a pulsed optical excitation is studied using the quantum trajectory theory. In materials with both TR symmetry and optical selection rules, the Faraday rotation angle of the optical emission delayed by multiples of the THz period is shown to be exactly equal to the Berry phase of the stationary quantum trajectory. In Sec. \ref{main}, we show the valley contrasted optical selection rules of the biased bilayer graphene, which makes it a good system for the observation of the Berry phases of quantum trajectories. Furthermore, calculation shows that the bilayer graphene has a Berry curvature much lager than that in monolayer MoS$_2$, which leads to a much larger Berry phase of the stationary trajectory and in turn much larger Faraday rotation. The giant Faraday rotation is verified by numerical simulations. Sec. \ref{conclusion} concludes this paper.

\section{Model and formalism}
\label{genernal}
We consider a semiconductor under an elliptically polarized THz field
\begin{equation}\label{Ft}
{\mathbf F}\left( t \right) = F \left( \cos \theta \cos \left(\omega t\right),\sin \theta \sin \left(\omega t\right), 0 \right),
\end{equation}
with $\omega$ much smaller than the energy gap of the material (so that the THz field does not induce interband tunneling). The Hamiltonian in the Bloch state representation $H \left({\mathbf k}\right)$ evolves adiabatically in the ${\mathbf k}$-space $H \left({\mathbf k}\right) \to H\left( {\tilde {{\mathbf k}}\left( t \right)} \right)$ under the driving of this field, with
\begin{align}
\notag \tilde {{\mathbf k}}\left( t \right)&= {\mathbf k} + e{\mathbf A}\left( t \right) \\
& = \left( {k_x  - k_0\cos\theta \sin \left(\omega t\right),k_y  + k_0\sin\theta \cos \left(\omega t\right)},k_z \right), \label{path}
\end{align}
where ${\mathbf A}\left( t \right)$ is the electromagnetism vector potential with ${\mathbf F}=  - \partial {\mathbf A}/\partial t$, and $k_0  = {eF}/\omega$. Then we study the interaction of this system with a weak optical laser that creates electron-hole pairs at the band edge. The interaction Hamiltonian is $\hat H_{\text{I}}=-\hat{{\mathbf P}}\cdot {\mathbf E}_{\text{I}}e^{-i\Omega t}+\text{h.c.}$. The interband polarization operator $\hat{{\mathbf P}}$ in the Bloch state representation is
\begin{equation}
\hat {{\mathbf P}} = \int {d{\mathbf k}} \hat e_{\mu, {\mathbf k}}^\dag \hat h_{\nu, -{\mathbf k}}^\dag {\mathbf{d}}_{\mu \nu,{\mathbf k}},\label{interdipole}
\end{equation}
where $\hat e$ and $\hat h$ are electron and hole operators, and the interband dipole moment ${{\mathbf d}}_{\mu \nu ,{{\mathbf k}}}$ is given by~\cite{Blount}
\begin{align}
{{\mathbf d}}_{\mu \nu ,{{\mathbf k}}}=\frac{{e\left\langle {+,\mu , {\mathbf k}} \right|i\nabla _{\mathbf k} H\left(\mathbf k\right)  \left| {-,\nu , {\mathbf k}} \right\rangle }}{{E^+_{\mathbf k}  - E^-_{\mathbf k} }}.
\end{align}
Here $+$ and $-$ denote the conduction and valence bands, respectively, $\mu$ and $\nu$ are the spin or pseudo-spin indices introduced to label degenerate bands,~\cite{note_index} and $\left| {\pm,\mu , {\mathbf k}} \right\rangle$ are the Bloch states of the conduction (valence) bands with eigenenergies $E_{{\mathbf k}}^{\pm}$. We assume that the initial state is the vacuum state $\left|G\right\rangle$ with empty conduction bands and filled valence bands. After excitation by an optical laser, the electron (hole) is driven into adiabatic evolution in the conduction (valence) band by the THz field, and thus obtains a geometric phase $\int {\left(\mathscr{A}_{\tilde {{\mathbf k}}}^ {\pm}\right)_{\mu\mu}\cdot d\tilde {{\mathbf k}}}$ in addition to the dynamical phase $-\int E_{\tilde{{\mathbf k}}}^{\pm}d\tau$, where $\left({\mathscr{A}}_{\tilde {{\mathbf k}}}^{\pm}\right)_{\mu \mu} = i\left\langle {\pm ,\mu , \tilde {{\mathbf k}}} \right|\nabla _{{\mathbf k}} \left| {\pm,\mu , \tilde {{\mathbf k}}} \right\rangle$ are the Berry connections with $\left| {\pm,\mu , \tilde {{\mathbf k}}} \right\rangle$ being the instantaneous eigenstates of $H\left( {\tilde {{\mathbf k}}\left( t \right)} \right)$ with instantaneous eigenenergies $E_{\tilde{{\mathbf k}}}^{\pm}$. In general, the Berry connection can be non-Abelian (i.e. $\left({\mathscr{A}}_{\tilde {{\mathbf k}}}^{\pm}\right)_{\mu \nu} = i\left\langle {\pm ,\mu , \tilde {{\mathbf k}}} \right|\nabla _{{\mathbf k}} \left| {\pm,\nu , \tilde {{\mathbf k}}} \right\rangle \ne 0$ for $\mu \ne \nu$) and the geometric phase factor becomes a unitary matrix $ \hat Te^{ i \int^t {{\mathscr{A}}_{\tilde{{\mathbf k}}\left( \tau \right)}^ \pm \cdot d\tilde {{\mathbf k}}\left( \tau \right)} }$, where $\hat T$ is the time-ordering operator. The electron and hole recombine at a later time, leading to optical emission modified by the geometric phase.


The linear optical response of THz field driven semiconductors has been formulated in Ref. \citenum{YF_MoS2}. If the THz field does not mix different spin or pseudo-spin states of the degenerate energy bands, the Berry connections are Abelian. The optical response is simplified to
\begin{align}
{\mathbf P}\left( t \right) = &\sum_{\mu} i\int_{ - \infty }^t {dt'} \int {d {\mathbf k}} {{\mathbf d}}_{\mu\mu,\tilde {{\mathbf k}} \left( t \right)}^* {{\mathbf d}}_{\mu\mu,\tilde {{\mathbf k}} \left( {t'} \right)}  \cdot {{\mathbf E}}_{\text{I}} \left(t'\right)\notag\\
& e^{ - i\int_{t'}^t {\varepsilon _{\tilde {{\mathbf k}}\left( \tau \right)} d\tau + i \int_{t'}^t { \left[{{\mathscr {A}}_{\tilde {{\mathbf k}}  \left( \tau \right)}  }\right]_{\mu\mu} \cdot d\tilde {{\mathbf k}} \left(\tau\right)}  - i\Omega t'} } , \label{AbelQT}
\end{align}
where $\varepsilon _{\tilde {{\mathbf k}}}  = E^{+}_{\tilde {{\mathbf k}}}  - E^{-}_{\tilde {{\mathbf k}}}$ is the energy of the electron-hole pair, and ${\mathscr{A}}_{\tilde{\mathbf k}}={\mathscr{A}}^+_{\tilde{\mathbf k}}- {\mathscr{A}}^-_{\tilde{\mathbf k}}$ is the combined Berry connection of the electron-hole pair.

Now we consider a system that has TR symmetry and nontrivial Berry curvatures in its energy bands. If one state of the Kramers pair is denoted as the pseudospin state $\Uparrow$ and the other as $\Downarrow$, we have the relations
\begin{equation}
{{\mathbf d}}_{ \Uparrow  \Uparrow ,{\mathbf k}}  = {{\mathbf d}}_{ \Downarrow  \Downarrow , - {\mathbf k}}^* :={{\mathbf d}}_{{\mathbf k}},\ \left({\mathscr{A}}_{{\mathbf k}}\right) _{\Uparrow  \Uparrow}  = \left({\mathscr{A}}_{ - {\mathbf k}}\right)_{\Downarrow \Downarrow}^*:={\mathscr{A}}_{{\mathbf k}}, \label{TRrelation}
\end{equation}
according to the TR symmetry. We apply a short optical laser pulse to the system at time $t' = 0$, with the duration much shorter than the THz period $T=2\pi/\omega$. The pulse can be approximated by a $\delta$-pulse ${{\mathbf E}}_{\text{I}}\left(t'\right)\approx {\mathbf E}\delta\left(t'\right)$.  After an integer multiple of the THz period $t_n=nT$, the electron (hole) undergoes a cyclic evolution in the conduction (valence) band and the geometric phase becomes the gauge invariant Berry phase.~\cite{Berry} Using equation (\ref{TRrelation}), we get
\begin{equation}
\phi_B^{\left(n\right)}\left({\mathbf k}\right)=\phi^{\left(n\right)}_{B,\Uparrow\Uparrow}\left({\mathbf k}\right)
=-\phi^{\left(n\right)}_{B,\Downarrow\Downarrow}\left({\mathbf k}\right)
=\int_{0}^{t_n} {{\mathscr{A}}_{\tilde {{\mathbf k}}}  \cdot d\tilde {{\mathbf k}}},\label{Opposite_phase}
\end{equation}
i.e. the two TR related states have opposite Berry phases. Then the response at $t_n$ is simplified as
\begin{align}
{\mathbf P}\left( t_n \right) = & i\int {d{\mathbf k}} e^{ { - i\int_{0}^{t_n} {\varepsilon _{\tilde {{\mathbf k}}\left( \tau \right)} d\tau}  +i \phi_B^{\left(n\right)}\left({\mathbf k}\right)  } } {{\mathbf d}}_{+\tilde {{\mathbf k}}\left( 0 \right)}^* {{\mathbf d}}_{+\tilde {{\mathbf k}}\left( 0 \right)}  \cdot { {\mathbf E}} \ \notag\\
+ & i\int {d{\mathbf k}} e^{ { - i\int_{0}^{t_n} {\varepsilon _{\tilde {{\mathbf k}}\left( \tau \right)} d\tau}  -i \phi_B^{\left(n\right)}\left({\mathbf k}\right)  } } {{\mathbf d}}_{-\tilde {{\mathbf k}}\left( 0 \right)}{{\mathbf d}}^*_{-\tilde {{\mathbf k}}\left( 0 \right)}  \cdot { {\mathbf E}}. \label{RespN}
\end{align}
We see that ${\mathbf P}\left( t_n \right)$ is given by the interference between two responses with the same dynamical phase but opposite Berry phases, which, as shown below, results in a Faraday rotation of the optical emission.

The main consequence of equation~(\ref{RespN}) can be understood using the stationary phase formalism (or the quantum trajectory theory).~\cite{HSG_RBL,HHG_QT,YF_MoS2} In the path integral, the electron-hole pair moves along all possible trajectories when driven by the THz field, with the phase given by the action $S\left({\mathbf k}\right)= \int_{0}^{t_n} \varepsilon _{\tilde {{\mathbf k}}\left( \tau \right)} d\tau$. The Berry phase is dropped from this action that determines the electron-hole motion, since it is generally much smaller than the dynamical phase (see Fig.~\ref{saddle_Fig}). As the THz field is strong, the motion amplitude of the electron-hole pair is much larger than the wavepacket diffusion range. Thus the response is dominated by the stationary phase points of the action (i.e. trajectories that satisfy the least action condition)
\begin{equation}
\nabla_{{\mathbf k}}S\left({\mathbf k}\right)= \int_{0}^{t_n} {\mathbf v}_{\tilde{{\mathbf k}}} d\tau=0, \label{saddle}
\end{equation}
where ${{\mathbf v}}_{\tilde{{\mathbf k}}}=\nabla_{{\mathbf k}}{\varepsilon}_{\tilde{{\mathbf k}}}$ is the semiclassical velocity of the electron-hole pair. The stationary phase condition in Eq. (\ref{saddle}) means the electron under the acceleration by the THz field returns to the hole after $nT$ for recombination. Then the Berry phase is determined mainly by the stationary trajectory:
\begin{align}
{\mathbf P}\left( t_n \right) = & e^{+i \phi_B^{\left(n\right)}\left({\mathbf k}_s\right)} i\int {d{\mathbf k}} e^{ - i\int_{0}^{t_n} {\varepsilon _{\tilde {{\mathbf k}}\left( \tau \right)} d\tau}  } {{\mathbf d}}_{\tilde {{\mathbf k}}\left( 0 \right)}^* {{\mathbf d}}_{\tilde {{\mathbf k}}\left( 0 \right)}  \cdot { {\mathbf E}} \notag\\
+ & e^{-i \phi_B^{\left(n\right)}\left({\mathbf k}_s\right)} i\int {d{\mathbf k}} e^{ - i\int_{0}^{t_n} {\varepsilon _{\tilde {{\mathbf k}}\left( \tau \right)} d\tau}  } {{\mathbf d}}_{\tilde {{\mathbf k}}\left( 0 \right)}{{\mathbf d}}^*_{\tilde {{\mathbf k}}\left( 0 \right)}  \cdot { {\mathbf E}}. \label{RespQT}
\end{align}
where ${\mathbf d}_{-\tilde {\mathbf k}\left(0\right)}={\mathbf d}_{\tilde {\mathbf k}\left(0\right)}$ near the band edge has been used and ${\mathbf k}_s$ is the solution to equation (\ref{saddle}). The Berry phase $\phi^{\left(n\right)}_B\left({\mathbf k}_s\right)$ is accumulated along the stationary trajectory. When multiple solutions exist, the response is given by the interference of all possible stationary trajectories. The optical selection rule near the Dirac points in bilayer graphene is such that the interband dipole moment ${{\mathbf d}}_{{\mathbf k}} \approx d_{cv,{\mathbf k}}\left( {{\mathbf e}}_x  - i{{\mathbf e}}_y \right)/\sqrt 2$, i.e. the optical transitions at valley $\Uparrow$ ($\Downarrow$) is coupled to the $\sigma+$ ($\sigma-$) polarized light (see Sec.~\ref{main}). The linear susceptibilities for the $\sigma_{\pm}$-polarized lights are respectively
\begin{equation}
\chi_{\pm \pm}\left( t_n \right)= e^{\pm i \phi_B^{\left(n\right)}\left({\mathbf k}_s\right)} i\int {d{\mathbf k}} e^{ - i\int_{0}^{t_n} {\varepsilon _{\tilde {{\mathbf k}}\left( \tau \right)} d\tau}  } \left|d_{cv,\tilde {{\mathbf k}}\left( 0 \right)}\right|^2.
\end{equation}
If the excitation light is linearly polarized in the $x$-$y$ plane, the Faraday rotation of the optical emission at $t_n$ is exactly given by the Berry phase ${\phi}_{FR}\left(t_n\right) = \phi_B^{\left(n\right)}\left({\mathbf k}_s\right)$.

The Faraday rotation caused by the elliptically polarized THz field can be intuitively understood as following.~\cite{YF_MoS2} The linearly polarized optical laser is a superposition of two opposite circular polarizations, which cause respective transitions of the two Kramers states. The electron-hole pair created by the optical pulse is in a superposition of the Kramers states $\left|  \Uparrow  \right\rangle  + \left|  \Downarrow  \right\rangle$. After the cyclic evolution under the THz field, the $\Uparrow$ and $\Downarrow$ states obtain the same dynamical phase $\phi_D$ and opposite Berry phases $\pm\phi_B$. Thus the final state of the electron-hole pair is $e^{i\phi _D } \left( {e^{i\phi _B} \left|  \Uparrow  \right\rangle  + e^{ - i\phi _B } \left|  \Downarrow  \right\rangle } \right)$, which gives emission with linear polarization rotated by an angle $\phi_B$.

From the preceding discussion it is clear that the theory can be applied to any material with TR symmetry, optical selection rules and nontrivial Berry phases, such as the topological insulators,~\cite{SCZhang,Kane} monolayer $\rm{MoS}_2$ and other group-VI dichalcogenides~\cite{MOS2_Mak1,MOS2,MOS2_Zeng,MOS2_Mak,MOS2_Cao} and bilayer graphene.~\cite{BilayerG,Graphene_RMP,Bilayer_Tunable,Bilayer_ARPES} In Ref. \citenum{YF_MoS2}, the monolayer $\rm{MoS}_2$ was studied and the result -- Faraday rotation equals the Berry phase of the stationary trajectory -- was verified by numerical calculation. In Sec.~\ref{main}, we will investigate the bilayer graphene. It has about 100 times larger Berry curvature at the band edges than the monolayer $\rm{MoS}_2$ does, which leads to a giant Faraday rotation of the optical emission.

\section{Berry phase induced Faraday rotation in bilayer graphene}
\label{main}

In bilayer graphene with Bernal (A-B) stacking, the band structure is well described by the tight-binding Hamiltonian with an intralayer nearest-neighbor hopping $t\approx 3 \text{eV}$, an interlayer nearest-neighbor hopping $\gamma \approx 0.4 \text{eV}$, and an interlayer bias $2\Delta$,~\cite{BilayerG,Bilayer_Tunable,Graphene_RMP}
\begin{equation}
H\left( \mathbf K \right) = \left( {\begin{array}{*{20}c}
   \Delta & {f\left( \mathbf K \right)} & 0 & 0  \\
   {f^* \left( \mathbf K \right)} & \Delta & {\gamma } & 0  \\
   0 & {\gamma } & { - \Delta} & {f\left( \mathbf K \right)}  \\
   0 & 0 & {f^* \left( \mathbf K \right)} & { - \Delta}  \\
\end{array}} \right).
\end{equation}
Here $f\left(\mathbf K\right)=-t\left(e^{i{\mathbf K}\cdot \mathbf{b}_1}+e^{i{\mathbf K}\cdot \mathbf{b}_2}+e^{i{\mathbf K}\cdot \mathbf{b}_3}\right)$, where $\mathbf{b}_{1,2} = \left(\frac{a}{2\sqrt 3}, \pm \frac{a}{2}\right)$ and $\mathbf{b}_{3} = \left(-\frac{a}{\sqrt 3}, 0\right)$ with $a \approx 2.46 \AA$ being the lattice constant. Expanding the momentum near the two Dirac points $\mathbf{K}_{\pm} = \left(0,\pm \frac{4\pi}{3a}\right)$, we have $f \left( {{\mathbf K}_ \pm   + \mathbf k} \right) = iv_f \left( {k_x  \mp ik_y } \right)$, where $v_f=\frac{\sqrt 3 }{2}at \, (\approx 10^6 \text{m/s})$. The states at $K_\pm$ valleys are related by TR operation and the corresponding Hamiltonians are given by
\begin{equation}
H_{\Uparrow/\Downarrow}\left( \mathbf k \right) = \left( {\begin{array}{*{20}c}
   \Delta & {iv_f k_\mp} & 0 & 0  \\
   {-iv_f k_\pm} & \Delta & {\gamma } & 0  \\
   0 & {\gamma } & { - \Delta} & {iv_f k_\mp}  \\
   0 & 0 & {-iv_f k_\pm} & { - \Delta}  \\
\end{array}} \right),
\end{equation}
where $k_\pm= k_x  \pm ik_y$ and the $K_\pm$ valley is denoted by the pseduo-spin notation $\Uparrow/\Downarrow$. In this paper, it is assumed that the valley coherence time in bilayer graphene is longer than the THz period and thus the intervalley scattering can be neglected. The calculated band structure (as shown in Fig.~\ref{Optical_sel}~(a)) reproduces the well-konwn Mexican hat structure.

The bilayer graphene has two conduction bands (positive energy bands) and two valence bands (negative energy bands). We assume that the Fermi level is kept in the energy gap while tuning the gap value.~\cite{Bilayer_Tunable} Hence the system is initially in the vacuum state $\left|G\right\rangle$ with empty conduction bands and filled valence bands. Since the optical laser is near resonant with the energy gap between the bottom conduction band and the top valence band, only the optical transitions between these two bands are considered, with $\left| {\pm,\mu , {\mathbf k}} \right\rangle$ ($\mu = \Uparrow, \Downarrow$) now denoting the corresponding Bloch states. Furthermore, it was shown in Ref. \citenum{YAO_selection} that there is a nearly perfect optical selection rule for the interband transitions between the two bands near the Dirac points, where the valley $K_\pm$ favors the $\sigma_\pm$ polarized transition, respectively. This can be seen from Fig. \ref{Optical_sel} (b), which plots the degree of circular polarization
\begin{equation}
\left[\eta\left({\mathbf k}\right)\right]_{\mu}  = \frac{{\left| d^+_{\mu\mu,\mathbf k} \right|^2  - \left| d^-_{\mu\mu,\mathbf k} \right|^2 }}{{\left| d^+_{\mu\mu,\mathbf k} \right|^2  + \left| d^-_{\mu\mu,\mathbf k} \right|^2 }},\label{deg_polar}
\end{equation}
with $d^\pm_{\mu\mu,\mathbf k} = {\mathbf d}_{\mu\mu,\mathbf k} \cdot \left({\mathbf e}_x  \pm i {\mathbf e}_y\right)/\sqrt 2$. Therefore, bilayer graphene serves as an ideal system for studying the Faraday rotation and Berry phases of quantum trajectories.

Now we calculate the combined Berry curvature of the bottom conduction band and the top valence band, defined as
\begin{equation}
\left(\Omega_{xy}\right)_{\mu\mu}= \left(\Omega^{+}_{xy}\right)_{\mu\mu}  - \left(\Omega^{-}_{xy}\right)_{\mu\mu},
\end{equation}
where
\begin{widetext}
\begin{align}
&\left(\Omega^{j}_{xy}\right)_{\mu\mu}= \partial _{k_x } \left({\mathscr A}^{j}_{k_y}\right)_{\mu\mu}  - \partial _{k_y } \left({\mathscr A}^{j}_{k_x}\right)_{\mu\mu} = i\sum\limits_{n \ne j} {\frac{{\left\langle {j,\mu ,\mathbf k} \right|\partial _{k_x } H_\mu\left(\mathbf k\right) \left| {n,\mu ,\mathbf k} \right\rangle \left\langle {n,\mu ,\mathbf k} \right|\partial _{k_y } H_\mu\left(\mathbf k\right) \left| {j,\mu ,\mathbf k} \right\rangle }}{{\left( {E_{\mathbf k}^j  - E_{\mathbf k}^n } \right)^2 }}} - \left( {x \leftrightarrow y} \right),
\end{align}
\end{widetext}
with $j=\pm$ and $\left| {n,\mu , {\mathbf k}} \right\rangle$ denoting the four eigenstates of $H_{\mu}\left( {\mathbf k} \right)$ with eigenenergy $E^{n}_{\mathbf k}$. Note that the Berry curvature is Abelian. The states at the two valleys have opposite Berry curvatures and thus give opposite Berry phases as pointed out in Sec.~\ref{genernal}. If there is no interlayer bias, the electron (hole) acquires a $\pm 2\pi$ Berry phase (or pseudospin winding number $\pm 2$) after traversing a path around the Dirac points ${\mathbf K}_{\pm}$.~\cite{Graphene_RMP,Pseudo_widing} When a bias $2\Delta = 0.3$ eV is applied, the distribution of $\Omega_{xy}$ is shown Fig.~\ref{Optical_sel} (c). The Berry curvature is concentrated in a small region enclosed by the band edge of the Mexican hatlike energy dispersion, being much larger than in monolayer MoS$_2$ ($\approx 22.3$ $\left(\pi/a\right)^{-2}$).~\cite{YF_MoS2,MOS2} Since the Berry phase equals the Berry curvature flux through the area enclosed by the path in the $\mathbf k$-space, the electron-hole pair in bilayer graphene acquires a much larger Berry phase during the cyclic evolution, which leads to a giant Faraday rotation of the optical emission.

\begin{figure}
\begin{center}
\includegraphics[width=\columnwidth]{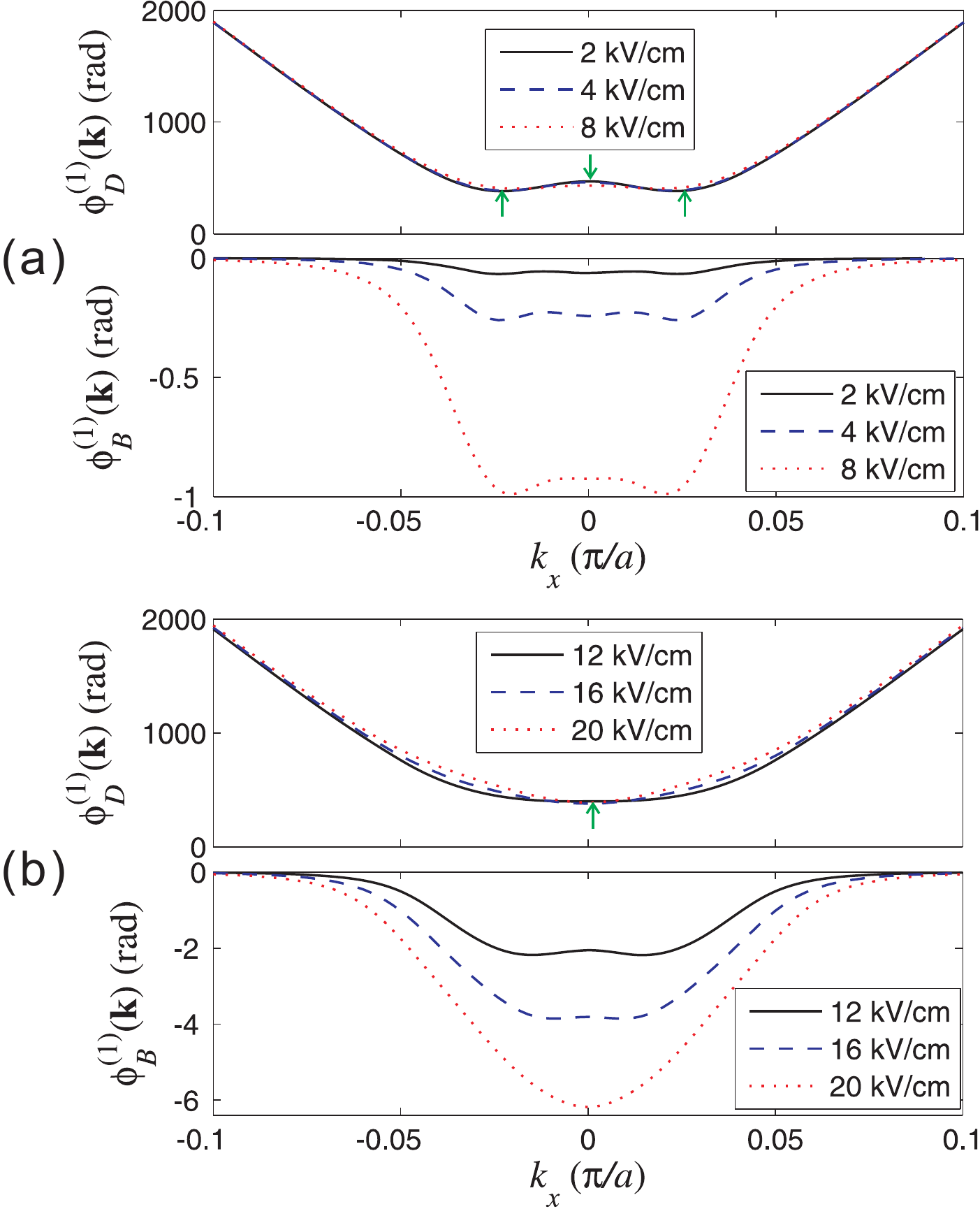}
\end{center}
\caption{(color online). The dynamical phase $\phi_D^{\left(1\right)}\left(\mathbf k\right)$ and Berry phase $\phi_B^{\left(1\right)}\left(\mathbf k\right)$ near the $K_+$ valley ($k_y=0$) over one THz period of evolution. Here we follow the notations in equations (\ref{TRrelation}) and (\ref{Opposite_phase}). The interlayer bias is $2\Delta=0.3$ eV, the frequency of the THz field is $\omega=4$ meV and the THz field is circularly polarized (i.e. $\theta=\pi/4$). In (a), the THz field strength $F = 2$, $4$ and $8$ $\text{kV/cm}$ (i.e., $k_0 = 0.0039$, $0.0078$ and $0.0157$ $\pi/a$), and the dynamical phase has two kinds of stationary phase points (indicated by the arrows) at the Dirac point and the bandedge. In (b), the THz field strength $F = 12$, $16$ and $20$ $\text{kV/cm}$ (i.e., $k_0 = 0.0235$, $0.0313$ and $0.0392$ $\pi/a$), and only the Dirac point is the stationary phase point (indicated by the arrow). } \label{saddle_Fig}
\end{figure}

Then we study the stationary trajectories, i.e. the stationary phase points of the dynamical phase $\phi_D^{\left(n\right)}\left(\mathbf k\right)= \int_{0}^{t_n} \varepsilon _{\tilde {{\mathbf k}}\left( \tau \right)} d\tau$, under the driving of the THz field. Figure \ref{saddle_Fig} plots the distribution of the dynamical phase $\phi_D^{\left(1\right)}\left(\mathbf k\right)$ over one THz period and the corresponding Berry phase $\phi_B^{\left(1\right)}\left(\mathbf k\right)$ at the $K_+$ valley for different filed strengths $F$. The dynamical phase is indeed much larger than the Berry phase. This justifies the approximation in Eq. (\ref{saddle}) for determining the quantum trajectories. When the THz field is weak (i.e. $k_0$, the amplitude of the quantum trajectory in $k$-space, is small as compared with the radius of the ``Mexican hat'' ring ($\approx 0.0233\pi/a$) around the Dirac point), we can expand $\varepsilon _{\tilde k\left( \tau  \right)}$ around $\mathbf k$ and get
\begin{align}
\phi _D^{\left( 1 \right)} \left( k \right) \approx \varepsilon _k T + \frac{1}{2}\frac{{\partial ^2 \varepsilon _k }}{{\partial k_x^2 }}k_0^2 T.
\end{align}
Hence the stationary phase points are exactly the extreme points of the energy band, i.e. the Dirac points and the
ring-shaped bandedges (Fig. \ref{saddle_Fig} (a)). Note that the trajectory around the Dirac point has a Berry phase close to that of the trajectory around the bandedge point (Fig.~\ref{saddle_Fig} (a)). Thus the Faraday rotation is well approximated by the Berry phase of the stationary trajectory around the Dirac point. When the THz field is strong enough (i.e. $k_0$ is comparable to the ``Mexican hat'' ring radius ($\approx 0.0233\pi/a$)), only the Dirac points are the stationary phase points (Fig. \ref{saddle_Fig} (b)). In this case, the Faraday rotation is also given by the Berry phase of the stationary trajectory centered at the Dirac point.

\begin{figure}[b]
\begin{center}
\includegraphics[width=\columnwidth]{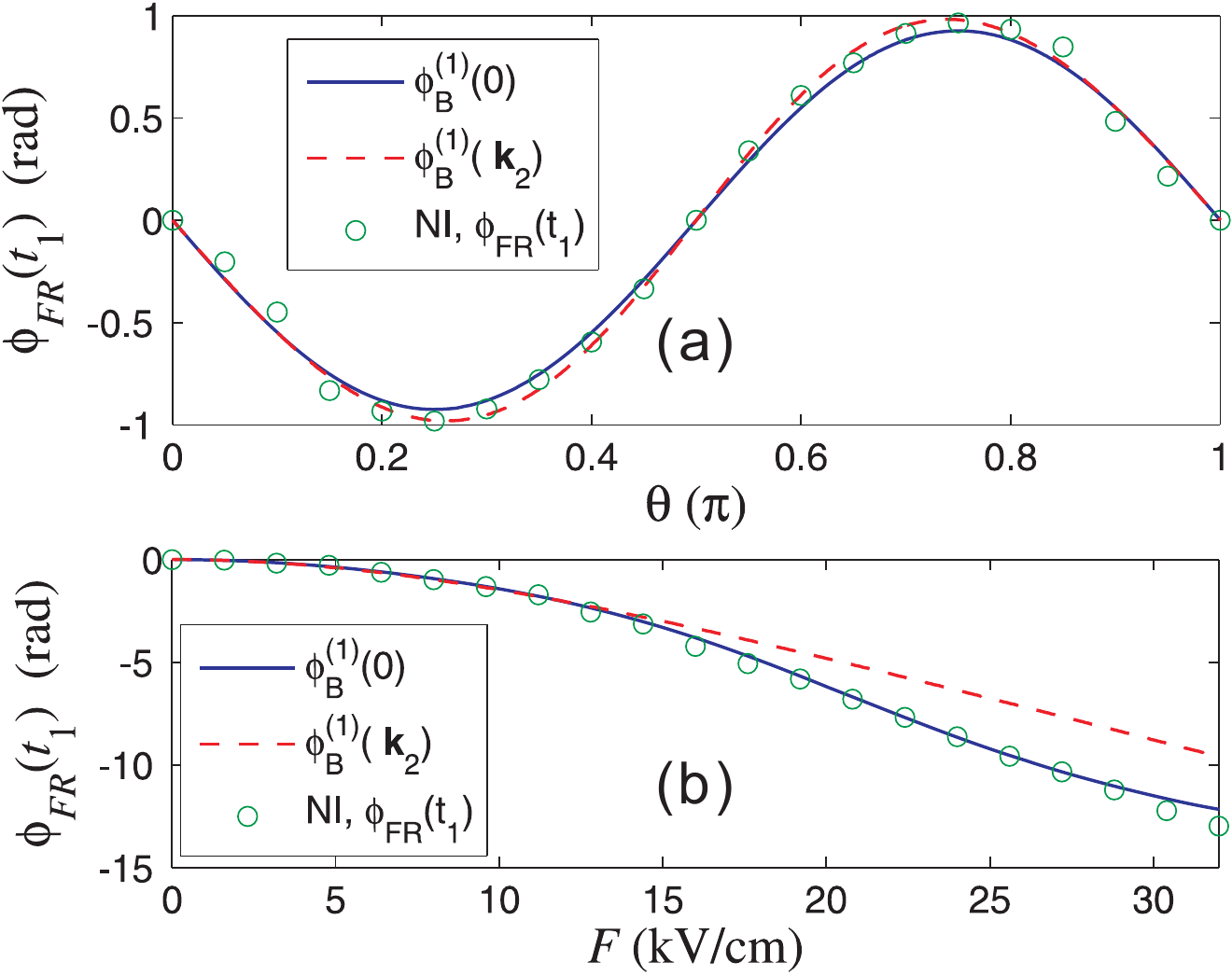}
\end{center}
\caption{(color online). Faraday rotation angle ${\phi}_{FR}$ of the optical emission at one THz period after pulse excitation, as a function of (a) the polarization ellipticity $\theta$ and (b) the strength $F$ of the THz field. The symbols show the numerical integration (NI) results and the lines are the Berry phases of the quantum trajectories centered at the stationary points (0 for the Dirac point, and ${\mathbf k}_2$ for the bandedge point with $k_y=0$). In (a), the THz field strength $F=8\ {\rm{kV/cm}}$ (i.e., $k_0=0.0157\pi/a$). In (b), the THz field is circularly polarized (i.e., $\theta=\pi/4$). The other parameters are the same as in Fig. 2.}
\label{Phir_fig}
\end{figure}

To verify the prediction of the quantum trajectory method (i.e. Faraday rotation equals the Berry phase of the stationary trajectory), we calculate the Faraday rotation directly through numerical integration of Eq.~(\ref{AbelQT}) for bilayer graphene and compare it with the Berry phase $\phi_B^{\left(n\right)}$ of the quantum trajectories. In the numerical calculation, the frequency of the THz field is $\omega=4$ meV and the optical pulse has the gaussian form $E{{\mathbf e}}_x \exp \left( { - i\Omega t  - t^2 /\delta t ^2 } \right)$, with $\Omega=2\Delta$ and $\omega \delta t  = 0.2$ ($\ll 2\pi$) (the optical laser spectrum width is $\sim 2/\delta t = $ 40 meV, much smaller than the valence/conduction band splitting). The results calculated for various THz field ellipticity $\theta$ and THz field strength $F$ are shown in Fig. \ref{Phir_fig}, which confirms that the Faraday rotation is well approximated by the Berry phase of the trajectory around the Dirac point. The Faraday rotation angle is about 100 times larger than in monolayer MoS$_2$,~\cite{YF_MoS2} consistent with the Berry curvature difference between the two materials. The Berry phase is given by the Berry curvature flux through the area element $\pi k_0^2 \sin\theta \cos\theta$ enclosed by the closed path in $\mathbf k$-space (equation (\ref{path})), which explains the approximate sinusoidal and parabolic dependence of the rotation angle on $\theta$ and $F$, respectively. Also as expected from the quantum trajectory analysis, the Berry phases around the bandedge points deviate appreciably from the numerical results when the THz field is strong enough (see Fig.~\ref{Phir_fig} (b)), which means the bandedge points fail to be the stationary phase points.

In the linear response regime, the giant Faraday rotation proposed is not affected by the recombination of the electron-hole pair. However, it requires the valley coherence time to be longer than the period of the THz field. Although the valley coherence time in bilayer graphene has not been determined to the best of our knowledge, the experiments for an analogous material (monolayer WSe$_2$) shows that the valley coherence time can be even longer than the recombination time of electron-hole pairs,~\cite{Valley_Coherence} which indicates the observation of the predicted giant Faraday rotation is promising. On the other hand, the Farday rotation can also be employed to study the intervalley coherence.

\section{Conclusions}
\label{conclusion}

In summary, the Berry phase dependent theory of THz extreme nonlinear optics developed in Ref. \citenum{YF_MoS2} is applied to the biased bilayer graphene. The Faraday rotation angle of the emission light delayed by integer multiples of the THz period is the Berry phase of the quantum trajectory. As the bilayer graphene has large Berry curvatures near the Dirac points, the Berry phase of the stationary trajectory is large, which leads to a giant Faraday rotation. The predictions by the quantum trajectory analysis are confirmed by numerical simulations. This result provides an opportunity to use bilayer graphene and relatively low-power THz lasers for ultrafast electro-optical devices.

\begin{acknowledgements}
We thank Mark S. Sherwin for stimulating discussions and helpful comments. This work is supported by Hong Kong RGC/GRF 401512.
\end{acknowledgements}


\begin{thebibliography}{26}
\expandafter\ifx\csname natexlab\endcsname\relax\def\natexlab#1{#1}\fi
\expandafter\ifx\csname bibnamefont\endcsname\relax
  \def\bibnamefont#1{#1}\fi
\expandafter\ifx\csname bibfnamefont\endcsname\relax
  \def\bibfnamefont#1{#1}\fi
\expandafter\ifx\csname citenamefont\endcsname\relax
  \def\citenamefont#1{#1}\fi
\expandafter\ifx\csname url\endcsname\relax
  \def\url#1{\texttt{#1}}\fi
\expandafter\ifx\csname urlprefix\endcsname\relax\def\urlprefix{URL }\fi
\providecommand{\bibinfo}[2]{#2}
\providecommand{\eprint}[2][]{\url{#2}}

\bibitem[{\citenamefont{Liu and Zhu}(2007)}]{HSG_RBL}
\bibinfo{author}{\bibfnamefont{R.-B.} \bibnamefont{Liu}} \bibnamefont{and}
  \bibinfo{author}{\bibfnamefont{B.-F.} \bibnamefont{Zhu}}, in
  \emph{\bibinfo{booktitle}{Physics of Semiconductors: 28th International
  Conference on the Physics of Semiconductors - ICPS 2006}}, edited by
  \bibinfo{editor}{\bibfnamefont{W.}~\bibnamefont{Jantsch}} \bibnamefont{and}
  \bibinfo{editor}{\bibfnamefont{F.}~\bibnamefont{Sch$\ddot{\rm a}$ffler}}
  (\bibinfo{organization}{AIP}, \bibinfo{address}{New York},
  \bibinfo{year}{2007}), vol. \bibinfo{volume}{893} of
  \emph{\bibinfo{series}{AIP Conf. Proc.}}, pp. \bibinfo{pages}{1455--1456}.

\bibitem[{\citenamefont{Zaks et~al.}(2012)\citenamefont{Zaks, Liu, and
  Sherwin}}]{Zaks_Liu}
\bibinfo{author}{\bibfnamefont{B.}~\bibnamefont{Zaks}},
  \bibinfo{author}{\bibfnamefont{R.~B.} \bibnamefont{Liu}}, \bibnamefont{and}
  \bibinfo{author}{\bibfnamefont{M.~S.} \bibnamefont{Sherwin}},
  \bibinfo{journal}{Nature} \textbf{\bibinfo{volume}{483}},
  \bibinfo{pages}{580} (\bibinfo{year}{2012}).

\bibitem[{\citenamefont{Krause et~al.}(1992)\citenamefont{Krause, Schafer, and
  Kulander}}]{HHG_Krause}
\bibinfo{author}{\bibfnamefont{J.~L.} \bibnamefont{Krause}},
  \bibinfo{author}{\bibfnamefont{K.~J.} \bibnamefont{Schafer}},
  \bibnamefont{and} \bibinfo{author}{\bibfnamefont{K.~C.}
  \bibnamefont{Kulander}}, \bibinfo{journal}{Phys. Rev. Lett.}
  \textbf{\bibinfo{volume}{68}}, \bibinfo{pages}{3535} (\bibinfo{year}{1992}).

\bibitem[{\citenamefont{Corkum}(1993)}]{HHG_Corkum}
\bibinfo{author}{\bibfnamefont{P.~B.} \bibnamefont{Corkum}},
  \bibinfo{journal}{Phys. Rev. Lett.} \textbf{\bibinfo{volume}{71}},
  \bibinfo{pages}{1994} (\bibinfo{year}{1993}).

\bibitem[{\citenamefont{Lewenstein et~al.}(1994)\citenamefont{Lewenstein,
  Balcou, Ivanov, L'Huillier, and Corkum}}]{HHG_QT}
\bibinfo{author}{\bibfnamefont{M.}~\bibnamefont{Lewenstein}},
  \bibinfo{author}{\bibfnamefont{P.}~\bibnamefont{Balcou}},
  \bibinfo{author}{\bibfnamefont{M.~Y.} \bibnamefont{Ivanov}},
  \bibinfo{author}{\bibfnamefont{A.}~\bibnamefont{L'Huillier}},
  \bibnamefont{and} \bibinfo{author}{\bibfnamefont{P.~B.}
  \bibnamefont{Corkum}}, \bibinfo{journal}{Phys. Rev. A}
  \textbf{\bibinfo{volume}{49}}, \bibinfo{pages}{2117} (\bibinfo{year}{1994}).

\bibitem[{\citenamefont{Corkum}(2011)}]{HHG_Phytoday}
\bibinfo{author}{\bibfnamefont{P.~B.} \bibnamefont{Corkum}},
  \bibinfo{journal}{Phys. Today} \textbf{\bibinfo{volume}{64}},
  \bibinfo{pages}{36} (\bibinfo{year}{2011}).

\bibitem[{\citenamefont{Yang and Liu}()}]{YF_MoS2}
\bibinfo{author}{\bibfnamefont{F.}~\bibnamefont{Yang}} \bibnamefont{and}
  \bibinfo{author}{\bibfnamefont{R.-B.} \bibnamefont{Liu}},
  \bibinfo{note}{arXiv:1211.3021}.

\bibitem[{\citenamefont{Ashcroft and Mermin}(1976)}]{Solid_AM}
\bibinfo{author}{\bibfnamefont{N.~W.} \bibnamefont{Ashcroft}} \bibnamefont{and}
  \bibinfo{author}{\bibfnamefont{N.~D.} \bibnamefont{Mermin}},
  \emph{\bibinfo{title}{Solid State Physics}} (\bibinfo{publisher}{Saunders},
  \bibinfo{address}{Philadelphia}, \bibinfo{year}{1976}).

\bibitem[{\citenamefont{Berry}(1984)}]{Berry}
\bibinfo{author}{\bibfnamefont{M.~V.} \bibnamefont{Berry}},
  \bibinfo{journal}{Proc. R. Soc. Lond., Ser. A}
  \textbf{\bibinfo{volume}{392}}, \bibinfo{pages}{45} (\bibinfo{year}{1984}).

\bibitem[{\citenamefont{Qi and Zhang}(2011)}]{SCZhang}
\bibinfo{author}{\bibfnamefont{X.-L.} \bibnamefont{Qi}} \bibnamefont{and}
  \bibinfo{author}{\bibfnamefont{S.-C.} \bibnamefont{Zhang}},
  \bibinfo{journal}{Rev. Mod. Phys.} \textbf{\bibinfo{volume}{83}},
  \bibinfo{pages}{1057} (\bibinfo{year}{2011}).

\bibitem[{\citenamefont{Hasan and Kane}(2010)}]{Kane}
\bibinfo{author}{\bibfnamefont{M.~Z.} \bibnamefont{Hasan}} \bibnamefont{and}
  \bibinfo{author}{\bibfnamefont{C.~L.} \bibnamefont{Kane}},
  \bibinfo{journal}{Rev. Mod. Phys.} \textbf{\bibinfo{volume}{82}},
  \bibinfo{pages}{3045} (\bibinfo{year}{2010}).

\bibitem[{\citenamefont{Xie et~al.}()\citenamefont{Xie, Zhu, and Liu}}]{XTXie}
\bibinfo{author}{\bibfnamefont{X.-T.} \bibnamefont{Xie}},
  \bibinfo{author}{\bibfnamefont{B.-F.} \bibnamefont{Zhu}}, \bibnamefont{and}
  \bibinfo{author}{\bibfnamefont{R.-B.} \bibnamefont{Liu}},
  \bibinfo{note}{arXiv:1305.5611}.

\bibitem[{\citenamefont{McCann and Fal'ko}(2006)}]{BilayerG}
\bibinfo{author}{\bibfnamefont{E.}~\bibnamefont{McCann}} \bibnamefont{and}
  \bibinfo{author}{\bibfnamefont{V.~I.} \bibnamefont{Fal'ko}},
  \bibinfo{journal}{Phys. Rev. Lett.} \textbf{\bibinfo{volume}{96}},
  \bibinfo{pages}{086805} (\bibinfo{year}{2006}).

\bibitem[{\citenamefont{Castro~Neto et~al.}(2009)\citenamefont{Castro~Neto,
  Guinea, Peres, Novoselov, and Geim}}]{Graphene_RMP}
\bibinfo{author}{\bibfnamefont{A.~H.} \bibnamefont{Castro~Neto}},
  \bibinfo{author}{\bibfnamefont{F.}~\bibnamefont{Guinea}},
  \bibinfo{author}{\bibfnamefont{N.~M.~R.} \bibnamefont{Peres}},
  \bibinfo{author}{\bibfnamefont{K.~S.} \bibnamefont{Novoselov}},
  \bibnamefont{and} \bibinfo{author}{\bibfnamefont{A.~K.} \bibnamefont{Geim}},
  \bibinfo{journal}{Rev. Mod. Phys.} \textbf{\bibinfo{volume}{81}},
  \bibinfo{pages}{109} (\bibinfo{year}{2009}).

\bibitem[{\citenamefont{Castro et~al.}(2007)\citenamefont{Castro, Novoselov,
  Morozov, Peres, dos Santos, Nilsson, Guinea, Geim, and
  Neto}}]{Bilayer_Tunable}
\bibinfo{author}{\bibfnamefont{E.~V.} \bibnamefont{Castro}},
  \bibinfo{author}{\bibfnamefont{K.~S.} \bibnamefont{Novoselov}},
  \bibinfo{author}{\bibfnamefont{S.~V.} \bibnamefont{Morozov}},
  \bibinfo{author}{\bibfnamefont{N.~M.~R.} \bibnamefont{Peres}},
  \bibinfo{author}{\bibfnamefont{J.~M.~B.~Lopes} \bibnamefont{dos Santos}},
  \bibinfo{author}{\bibfnamefont{J.}~\bibnamefont{Nilsson}},
  \bibinfo{author}{\bibfnamefont{F.}~\bibnamefont{Guinea}},
  \bibinfo{author}{\bibfnamefont{A.~K.} \bibnamefont{Geim}}, \bibnamefont{and}
  \bibinfo{author}{\bibfnamefont{A.~H.} \bibnamefont{Castro Neto}},
  \bibinfo{journal}{Phys. Rev. Lett.} \textbf{\bibinfo{volume}{99}},
  \bibinfo{pages}{216802} (\bibinfo{year}{2007}).

\bibitem[{\citenamefont{Ohta et~al.}(2006)\citenamefont{Ohta, Bostwick,
  Seyller, Horn, and Rotenberg}}]{Bilayer_ARPES}
\bibinfo{author}{\bibfnamefont{T.}~\bibnamefont{Ohta}},
  \bibinfo{author}{\bibfnamefont{A.}~\bibnamefont{Bostwick}},
  \bibinfo{author}{\bibfnamefont{T.}~\bibnamefont{Seyller}},
  \bibinfo{author}{\bibfnamefont{K.}~\bibnamefont{Horn}}, \bibnamefont{and}
  \bibinfo{author}{\bibfnamefont{E.}~\bibnamefont{Rotenberg}},
  \bibinfo{journal}{Science} \textbf{\bibinfo{volume}{313}},
  \bibinfo{pages}{951} (\bibinfo{year}{2006}).

\bibitem[{\citenamefont{Xiao et~al.}(2012)\citenamefont{Xiao, Liu, Feng, Xu,
  and Yao}}]{MOS2}
\bibinfo{author}{\bibfnamefont{D.}~\bibnamefont{Xiao}},
  \bibinfo{author}{\bibfnamefont{G.-B.} \bibnamefont{Liu}},
  \bibinfo{author}{\bibfnamefont{W.}~\bibnamefont{Feng}},
  \bibinfo{author}{\bibfnamefont{X.}~\bibnamefont{Xu}}, \bibnamefont{and}
  \bibinfo{author}{\bibfnamefont{W.}~\bibnamefont{Yao}},
  \bibinfo{journal}{Phys. Rev. Lett.} \textbf{\bibinfo{volume}{108}},
  \bibinfo{pages}{196802} (\bibinfo{year}{2012}).

\bibitem[{\citenamefont{Yao et~al.}(2008)\citenamefont{Yao, Xiao, and
  Niu}}]{YAO_selection}
\bibinfo{author}{\bibfnamefont{W.}~\bibnamefont{Yao}},
  \bibinfo{author}{\bibfnamefont{D.}~\bibnamefont{Xiao}}, \bibnamefont{and}
  \bibinfo{author}{\bibfnamefont{Q.}~\bibnamefont{Niu}},
  \bibinfo{journal}{Phys. Rev. B} \textbf{\bibinfo{volume}{77}},
  \bibinfo{pages}{235406} (\bibinfo{year}{2008}).

\bibitem[{\citenamefont{Park and Marzari}(2011)}]{Pseudo_widing}
\bibinfo{author}{\bibfnamefont{C.-H.} \bibnamefont{Park}} \bibnamefont{and}
  \bibinfo{author}{\bibfnamefont{N.}~\bibnamefont{Marzari}},
  \bibinfo{journal}{Phys. Rev. B} \textbf{\bibinfo{volume}{84}},
  \bibinfo{pages}{205440} (\bibinfo{year}{2011}).

\bibitem[{\citenamefont{Blount}(1962)}]{Blount}
\bibinfo{author}{\bibfnamefont{E.~I.} \bibnamefont{Blount}}, in
  \emph{\bibinfo{booktitle}{Advances in Research and Applications}}, edited by
  \bibinfo{editor}{\bibfnamefont{F.}~\bibnamefont{Seitz}} \bibnamefont{and}
  \bibinfo{editor}{\bibfnamefont{D.}~\bibnamefont{Turnbull}}
  (\bibinfo{publisher}{Academic Press}, \bibinfo{address}{New York},
  \bibinfo{year}{1962}), vol.~\bibinfo{volume}{13} of
  \emph{\bibinfo{series}{Solid State Physics}}, pp. \bibinfo{pages}{305--373}.

\bibitem[{not()}]{note_index}
\bibinfo{note}{In the following discussions, we consider the time-reversal
  symmetric systems with two-fold Kramers degeneracy.}

\bibitem[{\citenamefont{Mak et~al.}(2010)\citenamefont{Mak, Lee, Hone, Shan,
  and Heinz}}]{MOS2_Mak1}
\bibinfo{author}{\bibfnamefont{K.~F.} \bibnamefont{Mak}},
  \bibinfo{author}{\bibfnamefont{C.}~\bibnamefont{Lee}},
  \bibinfo{author}{\bibfnamefont{J.}~\bibnamefont{Hone}},
  \bibinfo{author}{\bibfnamefont{J.}~\bibnamefont{Shan}}, \bibnamefont{and}
  \bibinfo{author}{\bibfnamefont{T.~F.} \bibnamefont{Heinz}},
  \bibinfo{journal}{Phys. Rev. Lett.} \textbf{\bibinfo{volume}{105}},
  \bibinfo{pages}{136805} (\bibinfo{year}{2010}).

\bibitem[{\citenamefont{Zeng et~al.}(2012)\citenamefont{Zeng, Dai, Yao, Xiao,
  and Cui}}]{MOS2_Zeng}
\bibinfo{author}{\bibfnamefont{H.}~\bibnamefont{Zeng}},
  \bibinfo{author}{\bibfnamefont{J.}~\bibnamefont{Dai}},
  \bibinfo{author}{\bibfnamefont{W.}~\bibnamefont{Yao}},
  \bibinfo{author}{\bibfnamefont{D.}~\bibnamefont{Xiao}}, \bibnamefont{and}
  \bibinfo{author}{\bibfnamefont{X.}~\bibnamefont{Cui}},
  \bibinfo{journal}{Nature Nanotech.} \textbf{\bibinfo{volume}{7}},
  \bibinfo{pages}{490} (\bibinfo{year}{2012}).

\bibitem[{\citenamefont{Mak et~al.}(2012)\citenamefont{Mak, He, Shan, and
  Heinz}}]{MOS2_Mak}
\bibinfo{author}{\bibfnamefont{K.~F.} \bibnamefont{Mak}},
  \bibinfo{author}{\bibfnamefont{K.}~\bibnamefont{He}},
  \bibinfo{author}{\bibfnamefont{J.}~\bibnamefont{Shan}}, \bibnamefont{and}
  \bibinfo{author}{\bibfnamefont{T.~F.} \bibnamefont{Heinz}},
  \bibinfo{journal}{Nature Nanotech.} \textbf{\bibinfo{volume}{7}},
  \bibinfo{pages}{494} (\bibinfo{year}{2012}).

\bibitem[{\citenamefont{Cao et~al.}(2012)\citenamefont{Cao, Wang, Han, Ye, Zhu,
  Shi, Niu, Tan, Wang, Liu et~al.}}]{MOS2_Cao}
\bibinfo{author}{\bibfnamefont{T.}~\bibnamefont{Cao}},
  \bibinfo{author}{\bibfnamefont{G.}~\bibnamefont{Wang}},
  \bibinfo{author}{\bibfnamefont{W.}~\bibnamefont{Han}},
  \bibinfo{author}{\bibfnamefont{H.}~\bibnamefont{Ye}},
  \bibinfo{author}{\bibfnamefont{C.}~\bibnamefont{Zhu}},
  \bibinfo{author}{\bibfnamefont{J.}~\bibnamefont{Shi}},
  \bibinfo{author}{\bibfnamefont{Q.}~\bibnamefont{Niu}},
  \bibinfo{author}{\bibfnamefont{P.}~\bibnamefont{Tan}},
  \bibinfo{author}{\bibfnamefont{E.}~\bibnamefont{Wang}},
  \bibinfo{author}{\bibfnamefont{B.}~\bibnamefont{Liu}}, \bibnamefont{et~al.},
  \bibinfo{journal}{Nature Commun.} \textbf{\bibinfo{volume}{3}},
  \bibinfo{pages}{887} (\bibinfo{year}{2012}).

\bibitem[{\citenamefont{Jones et~al.}()\citenamefont{Jones, Yu, Ghimire, Wu,
  Aivazian, Ross, Zhao, Yan, Mandrus, Xiao et~al.}}]{Valley_Coherence}
\bibinfo{author}{\bibfnamefont{A.~M.} \bibnamefont{Jones}},
  \bibinfo{author}{\bibfnamefont{H.}~\bibnamefont{Yu}},
  \bibinfo{author}{\bibfnamefont{N.~J.} \bibnamefont{Ghimire}},
  \bibinfo{author}{\bibfnamefont{S.}~\bibnamefont{Wu}},
  \bibinfo{author}{\bibfnamefont{G.}~\bibnamefont{Aivazian}},
  \bibinfo{author}{\bibfnamefont{J.~S.} \bibnamefont{Ross}},
  \bibinfo{author}{\bibfnamefont{B.}~\bibnamefont{Zhao}},
  \bibinfo{author}{\bibfnamefont{J.}~\bibnamefont{Yan}},
  \bibinfo{author}{\bibfnamefont{D.~G.} \bibnamefont{Mandrus}},
  \bibinfo{author}{\bibfnamefont{D.}~\bibnamefont{Xiao}}, \bibnamefont{et~al.},
  \bibinfo{note}{arXiv:1303.5318}.

\end{thebibliography}
\newcommand{\noopsort}[1]{} \newcommand{\printfirst}[2]{#1}
  \newcommand{\singleletter}[1]{#1} \newcommand{\switchargs}[2]{#2#1}

\end{document}